\documentclass[conference]{IEEEtran}
\usepackage[pdftex]{graphicx}
\usepackage[usenames,dvipsnames,table,xcdraw]{xcolor}
\usepackage{array}
\usepackage{multirow}
\usepackage[normalem]{ulem}
\useunder{\uline}{\ul}{}
\usepackage{booktabs}
\usepackage{supertabular,booktabs}
\usepackage{longtable}
\usepackage{lscape}
\usepackage{cite}
\usepackage{url}
\usepackage{hyperref}
\usepackage{dirtytalk}
\usepackage{graphicx}
\usepackage{comment}
\usepackage{makecell}
\usepackage{svg}
\usepackage{amsmath}

\usepackage[caption=false]{subfig}
\graphicspath{{Images/}}
\usepackage[normalem]{ulem}
\useunder{\uline}{\ul}{}

\usepackage{tabularx}
\usepackage{balance}
\usepackage{nameref}

\usepackage[most]{tcolorbox}

\newtcolorbox{MyBox}{
  colback=gray!5!white,
  colframe=black!40!black,
  fonttitle=\bfseries,
  coltitle=black,
  sharp corners,
  boxrule=0.5pt,
  left=5pt,
  right=5pt,
  top=5pt,
  bottom=5pt,
  breakable
}

\begin{document}

\title{From Diverse Origins to a DEI Crisis: The Pushback Against Equity, Diversity, and Inclusion in Software Engineering}

\author{
\IEEEauthorblockN{Ronnie de Souza Santos}
\IEEEauthorblockA{University of Calgary\\
Calgary, AB, Canada \\
ronnie.desouzasantos@ucalgary.ca}
\and

\IEEEauthorblockN{Cleyton Magalhaes}
\IEEEauthorblockA{UFRPE\\
Recife, PE, Brazil\\
cleyton.vanut@ufrpe.br}
\and

\IEEEauthorblockN{Mairieli Wessel}
\IEEEauthorblockA{Radboud University\\
Nijmegen, Netherlands \\
mairieli.wessel@ru.nl}
\and

\IEEEauthorblockN{Ann Barcomb}
\IEEEauthorblockA{University of Calgary\\
Calgary, AB, Canada \\
ann.barcomb@ucalgary.ca}}


\IEEEtitleabstractindextext{%
\begin{abstract}
\textit{Background:}
Diversity, equity, and inclusion (DEI) are rooted in the very origins of software engineering and are shaped by the contributions of many individuals from underrepresented groups to the field. Yet today, DEI efforts in the industry face growing resistance, with companies retreating from visible commitments and pushing back on initiatives started only a few years ago.
\textit{Aims:} This study explores how the DEI backlash is unfolding in the software industry by investigating institutional changes, lived experiences, and the strategies used to sustain DEI practices. \textit{Method:} We conducted an exploratory case study using 59 publicly available Reddit posts authored by self-identified software professionals. Data were analyzed using reflexive thematic analysis. \textit{Results:} Our findings show that software companies are responding to the DEI backlash in varied ways, including re-structuring programs, scaling back investments, or quietly continuing efforts under new labels. Professionals reported a wide range of emotional responses, from anxiety and frustration to relief and happiness, shaped by identity, role, and organizational culture. Yet, despite the backlash, multiple forms of resistance and adaptation have emerged to protect inclusive practices in software engineering. \textit{Conclusions:} The DEI backlash is reshaping DEI in software engineering. While public messaging may soften or disappear, core DEI values persist in adapted forms. This study offers a new perspective on how inclusion is evolving under pressure and highlights the resilience of DEI in software environments.
\end{abstract}

}

\maketitle

\IEEEdisplaynontitleabstractindextext

\IEEEpeerreviewmaketitle

\section{Introduction}
\label{sec:introduction}
Diversity, equity, and inclusion (DEI), sometimes referred to as EDI, are interrelated concepts that aim to foster fair, respectful, and supportive environments for individuals from all backgrounds \cite{russen2023should}. Diversity refers to the presence of differences (e.g., race, gender, age), equity focuses on fairness by addressing unequal starting points, and inclusion emphasizes creating environments where everyone feels valued and heard \cite{russen2023should}. In software engineering, DEI plays a significant role in shaping team dynamics and the success of software systems \cite{rodriguez2021perceived, menezes2018diversity, kohl2022benefits, aleem2023practicing}.

Recent studies have demonstrated that DEI can lead to tangible benefits such as increased creativity, improved team effectiveness, stronger morale, greater engagement, and higher retention of developers \cite{verwijs2023double, aleem2023practicing, albusays2021diversity, rodriguez2021perceived, mason2024diversity}. At the same time, DEI can also introduce challenges that must be carefully managed, including the potential for relational conflict when inclusion and psychological safety are lacking, and the persistent emphasis on diversity over equity and inclusion, which limits the development of comprehensive strategies \cite{verwijs2023double, aleem2023practicing, rodriguez2021perceived, mason2024diversity}.

Even with the benefits of diversity, equity, and inclusion increasingly recognized across various domains, including software engineering, these principles have encountered growing resistance. This backlash has emerged in response to shifting public narratives, critiques of implementation, and debates about the scope and focus of DEI programs \cite{finn2023representation, mcgowan2025navigating, sitzmann2024don}. Critics argue that some initiatives are symbolic or unevenly applied, while others fear that DEI efforts may inadvertently create new forms of exclusion or division \cite{finn2023representation}. Still, others are motivated simply by political positioning or a lack of understanding about the historical and structural purposes of DEI work \cite{mcgowan2025navigating}. In the field of software engineering, this shifting landscape is prompting companies to adjust their internal policies, hiring practices, and training efforts, ultimately affecting how software professionals, particularly those from underrepresented groups, experience their workplace and their role within development teams \cite{dave2025, alfonseca2023, palmer2025}.

With diversity, equity, and inclusion efforts coming under increasing pressure, there is a growing need to understand how these shifts unfold within the software industry, particularly how they are being perceived and experienced by software professionals navigating this crisis. This study aims to investigate this problem through the following research questions:
\begin{itemize}
    \item \textbf{RQ1.} \textit{What changes caused by the DEI backlash have been observed and experienced in the software industry?}
    \item \textbf{RQ2.} \textit{How is the backlash against DEI affecting software professionals within the software industry?}
    \item \textbf{RQ3.} \textit{What strategies are currently being used in the software industry to protect and sustain DEI efforts in response to the backlash?}
\end{itemize}

From this introduction, this study is organized as follows. In Section~\ref{sec:background}, we present a literature review on the subject. Section~\ref{sec:method} describes our methodology. In Section~\ref{sec:findings}, we present our findings, which are discussed in Section~\ref{sec:discussion}, along with the implications and limitations of this study. Finally, Section~\ref{sec:conclusion} summarizes our contributions and final considerations.

\section{The DEI Backlash in the Software Industry} 
\label{sec:background}
Software engineering has always been shaped by individuals from diverse backgrounds, including many who broke ground despite being part of underrepresented groups. Ada Lovelace, considered the first computer programmer, Alan Turing, a gay mathematician whose work was foundational to AI, and Grace Hopper, who developed the first compiler, exemplify this legacy. Others, like Clarence ``Skip'' Ellis, the first African American with a PhD in computer science who made essential contributions to real-time collaborative systems, Peter Landin, an openly bisexual pioneer in programming language theory, and Christopher Strachey, who advanced formal methods, show that diversity is not a new goal in the field, but it is part of its foundation \cite{de2023diversity, de2024hidden, dodig2001history, mijwel2015history}.

Despite these historical roots, diversity remains a persistent challenge in software engineering. Systemic barriers continue to shape access and success, starting in academic settings and extending into industry. Underrepresented groups (including women, racialized individuals, LGBTQ+ people, and those with disabilities) face unequal access to support, mentorship, and advancement opportunities \cite{de2024diversity, ibe2018reflections, de2023lgbtqia+, hyrynsalmi2023diversity}. These inequities result in a narrowing pipeline from the university to the software industry, resulting in a lack of diverse perspectives in system design, impacting not only fairness but also innovation, usability, and security \cite{oliveira2024navigating, de2023benefits, zolduoarrati2021value, albusays2021diversity, adams2020diversity, rodriguez2021perceived}.

It is not enough that diversity in software engineering has remained a persistent challenge for decades, and the field is now facing a new and compounding crisis: a growing pushback against diversity, equity, and inclusion efforts. While attention to DEI gained momentum in the early 2020s, the scenario began to shift in 2023, when a wave of legal and political challenges started raising conflicts about the legitimacy and perceived fairness of diversity and inclusion initiatives. By 2024, several significant rollbacks were underway, and by 2025, these actions had escalated into a broader trend of retrenchment across the tech sector \cite{bryan2025, alfonseca2023, palmer2025, wong2025}. 

This shift is not only organizational but cultural. Internal communications and public statements reveal a move away from explicitly equity-focused strategies toward more generic expressions of inclusion and strategies aimed at minimizing controversy or legal exposure \cite{dave2025, wong2025}. Roles and departments specifically dedicated to DEI have been downsized or eliminated, and public commitments are increasingly being replaced by less targeted initiatives \cite{mcgowan2025navigating}. These changes have introduced uncertainty within software teams, particularly for professionals engaged in or benefited from earlier DEI initiatives.

Researchers caution that this backlash, while cyclical in nature, poses unique risks in the current landscape. Rather than signaling the resolution of equity challenges, the pullback may undermine long-term efforts to create more inclusive technical environments \cite{sitzmann2024don, finn2023representation}. In software engineering, the effects might be far-reaching, as it threatens the development of technologies that account for the needs of diverse users, weakens trust in design decisions, and narrows the perspectives contributing to critical systems \cite{albusays2021diversity, adams2020diversity, rodriguez2021perceived}.

\section{Method} 
\label{sec:method}
This work adopts an exploratory case study design to investigate how software professionals experience and respond to the growing backlash against diversity, equity, and inclusion initiatives in the software industry \cite{ralph2020empirical, runeson2009guidelines}.

\subsection{Research Design and Rationale}
Case studies are appropriate when studying contemporary phenomena in complex, real-world contexts, where boundaries between the phenomenon and its setting are unclear \cite{ralph2020empirical, runeson2009guidelines}. In our case, the DEI backlash cannot be easily separated from industry shifts, global political dynamics, and evolving workplace cultures.

This topic is also sensitive and politically charged. Software professionals may feel discouraged or even afraid to discuss their employers' DEI policies, especially given recent layoffs. Consequently, conducting interviews or identifying a specific organization as a case would have posed ethical and logistical challenges. Additionally, while survey-based methods can be used to assess broader populations, they are limited in their ability to capture rich personal narratives, and defining a representative sample of affected individuals across demographic groups is particularly difficult in this context.

Given these constraints, we selected Reddit as our data source. 
Our objective is not to generalize to all software professionals, but to explore how those participating in public discourse perceive and articulate their experiences with the DEI backlash by treating the Reddit posts as archival records of discourse, repurposed for qualitative analysis \cite{fisher2018research, nguyen2015internet}.

\subsection{The Case Site} 
Reddit is a popular social media platform organized into user-moderated communities called subreddits, each dedicated to specific topics, professions, or identity groups. These communities facilitate semi-anonymous, many-to-many conversations, where users post questions, share experiences, and comment in threaded discussions. Reddit is widely used among software professionals \cite{fang2023understanding}, and its public nature makes it well-suited for archival internet research \cite{nguyen2015internet}.

For this study, we defined our case site as a cluster of six subreddits: \textit{r/softwaredevelopment}, \textit{r/SoftwareEngineering}, \textit{r/womenintech}, \textit{r/AskLGBT}, \textit{r/Neurodivergent}, and \textit{r/cptsd\_bipoc}. We began by selecting subreddits focused on software development and engineering that demonstrated \textit{high activity levels and consistent engagement} (e.g., \textit{r/softwaredevelopment}, \textit{r/SoftwareEngineering}, and \textit{r/womenintech}). However, in order to reach a broader range of perspectives, particularly from professionals belonging to underrepresented groups, we expanded our selection to include identity-based subreddits where members often discuss work-related issues from lived experiences (e.g., \textit{r/AskLGBT}, \textit{r/Neurodivergent}, and \textit{r/cptsd\_bipoc}).

To investigate how software professionals were experiencing the DEI backlash, our initial plan was to identify and analyze existing posts on Reddit that already discussed the phenomenon. While we found general conversations about the DEI backlash across several subreddits, none addressed it in the context of the software industry, or offered detailed, experience-based accounts from software professionals themselves. Given this lack of contextual evidence, we opted to seed the discussion ourselves. To do so, we created an open-ended post in each subreddit to prompt voluntary narratives directly tied to our research questions:

\begin{MyBox}\textbf{Any Other Software Devs Here? How Are You Handling the Anti-DEI Wave?}
\small
Hey everyone, just curious—are there other software developers here? Lately, there’s been a lot of backlash against DEI efforts, and I’ve been wondering how others in the software industry are experiencing it, especially with big tech companies scaling back DEI initiatives.
\end{MyBox}

This strategy allowed us to directly reach software professionals and underrepresented individuals in tech communities while avoiding intrusive methods, aligning with case study conventions \cite{ralph2020empirical} and indirect observation approaches, allowing participants to articulate their experiences in their own words and on their own terms.

\subsection{Data Collection}
Data were collected during March 2025. After seeding the post in each subreddit, we manually screened all responses to identify relevant data. We included posts that:

\begin{itemize}
    \item Were made by individuals who claimed to work in the software industry;
    \item Addressed an experience, observation, or concern explicitly related to DEI practices or their rollback in the software industry;
    \item Demonstrated evidentiary value, meaning the post described a situation, reaction, or institutional change that could be interpreted within the broader theme of the DEI backlash.
\end{itemize}

Posts that were off-topic, vague, purely opinion-based without anchoring in experience, or unrelated to software contexts were excluded. The final dataset consisted of 59 unique comments and responses from the six subreddits. Each post was assigned an anonymized identifier for analysis (e.g., comment\_01, comment\_02). No usernames, timestamps, or metadata were retained.

\subsection{Data Analysis}

We conducted a reflexive thematic analysis \cite{cruzes2011recommended, clarke2017thematic} to identify how participants constructed meaning around the DEI backlash and to trace recurring patterns in their narratives. Two researchers worked independently on this task, and we did not use automated tools, e.g., all coding was conducted manually in shared documents to preserve the contextuality of each comment. The analysis involved the following steps:

\begin{itemize}
    \item \textbf{Initial familiarization}: Both researchers independently read the full dataset multiple times to identify early patterns and note potential codes.
    \item \textbf{Open coding}: Each researcher manually coded the dataset line by line, assigning descriptive labels to meaningful narrative segments. Coding was inductive, allowing for bottom-up construction of themes.
    \item \textbf{Code reconciliation}: Researchers met to review their codes, resolve discrepancies through discussion, and refine the themes collaboratively. A third researcher was available to adjudicate if needed.
    \item \textbf{Theme development}: Once the coding stabilized, we organized codes into candidate themes, reviewing for internal coherence and distinctiveness. Final themes were named and defined through iterative discussion.
\end{itemize}

Finally, saturation is a key principle in qualitative research, indicating that enough data has been collected to capture the main patterns in participants' experiences \cite{ralph2020empirical}. In our study, the amount of data depended on the willingness of participants on the platform to engage in the discussion, and although our dataset is small, it was rich in detail. Considering this limitation, we assessed saturation throughout the coding process and determined it had been reached when no new codes or meaningful themes were emerging from the data we obtained.

\subsection{Ethical Considerations}
This study involved an archival analysis of publicly available data from Reddit. Although user accounts were required to make the original posts, we only collected data after logging out, ensuring that we only accessed content available to the general public. Posts were collected without usernames, profile links, or any identifying information. Quotes used in reporting have been lightly edited for clarity and de-identified to protect anonymity. Finally, because the data were public, non-interactive, and presented minimal risk to participants, the study was reviewed and deemed exempt by the first author's institutional Research Ethics Board.

\section{Results}  \label{sec:findings}

Our dataset consists of 59 posts collected from five Reddit communities where members discussed the current backlash against diversity, equity, and inclusion in the software industry. The majority of the posts came from the subreddit \textit{r/womenintech}, accounting for 66.1\% (39/59) of the dataset. This was followed by \textit{r/SoftwareEngineering} with 16.9\% (10/59), \textit{r/softwaredevelopment} with 8.5\% (5/59), and \textit{r/AskLGBT} with 6.8\% (4/59). One post (1.7\%) was contributed from the community \textit{r/cptsd\_bipoc}. These varied sources offer insights into how the DEI backlash is perceived and experienced across different groups within the software development ecosystem.

Additionally, we observed notable differences in the tone and dynamics of the conversations depending on the community in which the posts appeared. In forums centered on underrepresented groups, such as \textit{r/womenintech}, \textit{r/AskLGBT}, and \textit{r/cptsd\_bipoc}, discussions were generally more progressive and approached the topic of DEI with empathy and openness. Even when users expressed disagreement, their exchanges tended to remain respectful and constructive, reflecting a shared commitment to thoughtful engagement with DEI-related issues. In contrast, the two general software engineering communities (\textit{r/SoftwareEngineering} and \textit{r/softwaredevelopment}) presented a different atmosphere. 

Several posts in these forums included negative views and attacks toward DEI and people from underrepresented groups, sometimes accompanied by discriminatory remarks or dismissive language. During the data collection process, we observed that many of these comments were deleted by the users shortly after posting, which could mean that users regretted engaging. As the posts were deleted, we were unable to identify their content or include them in the analysis.
Eventually, moderators locked both threads in the general software engineering communities within six hours of their creation, effectively ending further engagement. By contrast, all threads in the communities centered on underrepresented groups remained open for discussion, although no new interactions were observed after the end of the collection period.

\subsection{Observed Changes in the Software Industry Due to the DEI Backlash}

Our first set of findings reflected the observed or experienced changes in the software industry resulting from the DEI backlash. Out of the 59 collected posts, 21 described changes occurring within software organizations. Based on our thematic analysis we grouped them into four distinct categories that reflect how companies are responding to external pressures regarding DEI. The most frequently observed change was \textit{DEI Revisions} (18.6\%, 11 posts), followed by \textit{Job Cuts or Layoffs} (6.8\%, 4 posts), \textit{DEI Initiatives Cancelation} (5.1\%, 3 posts), and \textit{Reduction in DEI Investment} (5.1\%, 3 posts). The remaining 38 posts either stated that no changes had occurred so far or focused on other perspectives related to DEI. Below, we describe each category in detail.

\begin{itemize} 

 \item \textbf{DEI Revisions}: The evidence obtained from the posts demonstrates that some companies continued to support inclusive practices but removed explicit references to DEI, often rebranding programs to avoid criticism. In one specific example, the company replaced the acronym "DEI" with alternatives such as "IDEA" (Inclusion, Diversity, Equity, and Accessibility), or chose more neutral terms to maintain their programs. One user commented: \textit{``During a meeting for leaders across our various groups (queer folks, people of color, etc.), we were asked about 'rebranding' the term DEI externally since it had gotten a bad rep, while still pursuing the same mission of inclusiveness internally''} (Comment\_01). Another participant shared: \textit{``We still have inclusivity training and sensitivity training, but they are just called that and not DEI. Simple. Easy. Done''} (Comment\_30).

\item \textbf{Job Cuts or Layoffs}: In this theme, users reported job losses or reduced hiring prospects in environments where DEI was being challenged or deprioritized. Although not all layoffs were directly attributed to DEI backlash, some users linked job loss to DEI-related decisions, such as facing hiring freezes in environments that deprioritized diversity. One user wrote: \textit{``They created the bootcamp's first ever group applications made entirely by women. I got a PIP [Personal Improvement Plan] for doing this and was eventually fired''} (Comment\_21), referring to their decision to support an all-women project group. Another user reflected: \textit{``For me, I feel like the chances of me getting a job after my layoff 1.5 years ago is getting worse and worse''} (Comment\_07).

\item \textbf{DEI Initiatives Cancelation}: Another observed change includes more direct and comprehensive eliminations of DEI programs, including the disbanding of employee resource groups, removal of DEI-related language from public platforms, and elimination of minority groups' celebrations. These cancellations were sometimes tied to contracts or political pressure. One user noted: \textit{``They were required to sign something stating they didn’t practice DEI in order to keep the contracts. So they did and disbanded all the groups they had (women, lgbt, etc)''} (Comment\_14). Another wrote: \textit{``We had to remove all DEI keywords and phrases from our organization's websites and apps''} (Comment\_40).

\item \textbf{Reduction in DEI Investment}: We also identified that some companies did not cancel DEI programs outright but scaled back their visibility, funding, or external communications. Reductions often targeted advocacy efforts, marketing campaigns, or internal communication budgets. One user shared: \textit{``The internal advocate groups still exist but don’t get marketing money''} (Comment\_58). Another noted: \textit{``Our company released an internal news article about striking DEI programs''} (Comment\_53).

\end{itemize}

Taken together, these changes suggest a complex response to the DEI backlash in the software industry. While some companies have responded by scaling back, removing, or canceling their DEI-related efforts, others appear to be engaged in a form of quiet resistance, e.g., using rebranding to preserve the underlying goals of inclusion while adjusting the visibility of such programs, considering external and internal interaction.

\subsection{Emotional Responses and Lived Experiences of Software Professionals}

Our second set of findings focused on how professionals in the software industry are feeling in response to the ongoing DEI backlash. Out of the 59 collected posts, 24 described personal emotional responses or workplace experiences associated with the current climate. We grouped these responses into seven distinct categories that reflect a range of emotions and experiential states: \textit{Anxiety} (20.8\%, 5 posts), \textit{Frustration} (20.8\%, 5 posts), \textit{Relief} (20.8\%, 5 posts), \textit{Hopelessness} (12.5\%, 3 posts), \textit{Happiness} (12.5\%, 3 posts), \textit{Fear} (4.2\%, 1 post), and \textit{Uncertainty} (4.2\%, 1 post). These categories are not mutually exclusive and represent the dominant affective tone of each post. Below, we define each category and illustrate it with selected narratives.

\begin{itemize}
 
\item \textbf{Anxiety}: This category includes posts where participants explicitly mentioned being nervous, tense, or worried, particularly about the possibility of future discrimination or loss of job security. The anxiety expressed often stemmed from anticipation rather than concrete events. One participant explained: \textit{``So far I haven’t had any adverse action, but I’m tense and anxious''} (Comment\_34). Another user shared concerns about how DEI backlash might affect hiring practices: \textit{``I'm worried about interviewing now... I'm going to be interviewing as a woman in her 40s which is definitely a double whammy''} (Comment\_17).

\item \textbf{Frustration}: Posts about frustration often describe a disconnect between organizational values and the personal or professional identity of the participant. Frustration was also expressed when DEI efforts were perceived as shallow or when the culture no longer felt inclusive. One user described emotional exhaustion: \textit{``Feeling extremely burnt out and like nobody has my back at work''} (Comment\_23). Another shared: \textit{``It makes me want to quit tbh. It's hard to feel like tech aligns with my values anymore''} (Comment\_11).

\item \textbf{Relief}: This emotion was expressed by participants whose companies reaffirmed their DEI commitments or did not show signs of retreat. Relief emerged when participants felt protected by strong internal values or proactive leadership. One participant shared: \textit{``Our company did make it a point to emphasize that DEI efforts weren’t going anywhere, so that’s cool''} (Comment\_13). Another expressed satisfaction with recent messaging from leadership: \textit{``In the quarterly meeting, he talked about the importance of diversity and inclusion in our culture and mission statement (...) I feel a lot better about my company now''} (Comment\_16).

\item \textbf{Hopelessness}: This category captures emotional fatigue and resignation in response to repeated rejections, job insecurity, or emotional distress. These narratives often included language about giving up or feeling that things will be harder without DEI efforts. One user wrote: \textit{``I don’t even think it’s worth applying to anything because getting hit with 5-ish rejections a day is just miserable''} (Comment\_07). Another, after crying in front of their manager, summarized: \textit{``Badly.... Cried in front of my manager. Had a conversation with HR.... Shit's hard right now''} (Comment\_45).

\item \textbf{Happiness}: Some participants expressed happiness or satisfaction, because they disagreed with previous DEI practices. In these cases, the backlash was not experienced negatively, but rather as a return to something that the professionals felt more comfortable. One user shared: \textit{``I actually love it. Was so sick of being made to join female-focused affinity groups. Now I can just go to work and... work.''} (Comment\_29). Another expressed a desire for change against DEI: \textit{``It's a good thing, maybe finally we can return to the days of excellence''} (Comment\_47).

\item \textbf{Fear}: This post captured a sudden shift in the workplace environment following visible actions to erase DEI. The language used by the participant conveyed a sense of loss of safety and fear of retaliation: \textit{``For us they’ve gutted everything perceived as DEI related: from the office to people’s personal affects and scrubbed any digital footprint. Full on panic mode''} (Comment\_24). Fear in this context was closely tied to the speed and severity of the backlash.

\item \textbf{Uncertainty}: Finally, one post expressed vague apprehension about the future without reporting any concrete consequences. The participant noted: \textit{``I don’t know how long it’ll fly under their radar''} (Comment\_54), referring to their company’s keeping DEI initiatives amid pressure. This response reflected a sense of waiting and watching in a volatile environment. \end{itemize}

These findings reveal that the DEI backlash is not only a matter of institutional policy but a lived and deeply personal experience for many software professionals. While some find reassurance in strong leadership and consistency, others face growing emotional strain as they attempt to navigate ambiguous organizational climates. The feelings expressed, ranging from anxiety and frustration to relief and hopelessness, reflect a broad emotional spectrum shaped by individual roles, identities, and software development environments.

\subsection{Strategies to Protect or Sustain DEI Efforts in the Software Industry}

To complement our analysis of organizational changes and individual emotional responses, we also explored how software companies are responding to the DEI backlash through strategies of resistance and adaptation. Out of the 59 posts, 29 described some form of resistance or adaptation strategy. These were grouped into four categories: \textit{Organizational Resistance} (22.0\%, 13 posts), \textit{Rebranding} (20.3\%, 12 posts), \textit{Strategic Framing} (11.9\%, 7 posts), and \textit{Quiet Continuity} (10.2\%, 6 posts). These forms of adaptation reflect various ways that DEI efforts are being sustained or reshaped without being completely dismantled. Below, we define and describe each category with examples.

\begin{itemize} \item \textbf{Organizational Resistance}: Posts in this category described companies that explicitly reaffirmed their DEI values and continued their programs despite the broader backlash. This form of resistance involved visible and verbal commitments from leadership, maintenance of resource groups, and continued celebration of cultural events. One user noted: \textit{``ERGs [Employee Resource Groups] are if anything a bit more active than they used to be, plenty of pronouns in Slack and email signatures''} (Comment\_27). Another shared: \textit{``The big gigantic wall mural that says, ‘Diversity, Equity, Inclusion’ in 2-foot-high letters is still up. ERGs still meet''} (Comment\_26).

\item \textbf{Rebranding}: Rebranding refers to changing the terminology used to describe DEI efforts while continuing to implement them. This was the most tactical form of adaptation, aimed at appeasing critics without abandoning core practices. One user explained: \textit{``As someone who's been in branding for years, I can understand this move. It's at least better than completely abandoning the values''} (Comment\_02). Another shared: \textit{``I think the rebranding just gets [name removed] off your back and you can keep doing the same things. I would try to rebrand it as increasing team performance because it’s been proven that diversity does that''} (Comment\_06). Importantly, rebranding kept the structure of DEI initiatives intact but distanced them from politicized language.

\item \textbf{Strategic Framing}: This form involved adjusting both the language and presentation of DEI initiatives, sometimes in combination with operational changes, to align them with broader corporate goals or legal compliance. While rebranding changes only the terms, strategic framing includes modifying how the initiative is justified or implemented. One user described this process: \textit{``So they did and disbanded all the groups they had (women, lgbt, etc) and are `working on a solution.'\/''} (Comment\_14). Another noted: \textit{``The internal advocate groups still exist but dont get marketing money.''} (Comment\_58), reflecting a situation where DEI was paused or scaled back while the company looked for alternative approaches.

\item \textbf{Quiet Continuity}: Some users reported that DEI practices remained intact, but were not prominently advertised or labeled under the DEI banner. This subtle form of continuity often relied on geographic, organizational, or political positioning to avoid visibility. One user explained: \textit{``Nothing has changed. I moved to a blue city because I figured it would be better, and it really is''} (Comment\_28). Another shared: \textit{``We are continuing to hire a diverse workforce (``the right thing''), because we are too small for the anti-DEI witch-hunters to notice us''} (Comment\_59). \end{itemize}

These findings demonstrate that while the DEI backlash has changed the public visibility of diversity initiatives in the software industry, it has also given rise to a diverse set of responses aimed at sustaining inclusive practices. Whether through explicit resistance, quiet continuity, rebranding, or strategic repositioning, these strategies illustrate that, to some extent, support for DEI has not disappeared; it has adapted.

\section{Discussion} \label{sec:discussion}
We organize our discussion in three parts. First, we answer the research questions that guided our study. Second, we compare our results to related literature. Finally, we discuss the broader implications of our findings.

\subsection{Answering the Research Questions}

Each of our research questions offers a different perspective to understand the DEI backlash in the software industry. Hence, rather than treating the phenomenon a single issue, we explored it from three angles: institutional change, personal experience, and strategic adaptation. We can answer the questions as follows.

\noindent \textbf{RQ1. What changes caused by the DEI backlash have been observed and experienced in the software industry?}
The changes reported by participants are not limited to outright elimination of DEI programs; rather, they reflect a spectrum of responses, from visible dismantling to strategic adaptations. These organizational shifts reveal that the backlash in the software industry is less about ending DEI altogether and more about renegotiating how it is expressed and operationalized. In some cases, 
companies are making calculated moves to retain the spirit of DEI while minimizing exposure to avoid political and public controversies. In others, however, symbolic gestures give way to substantive rollbacks, leaving software engineers uncertain about the values their organizations uphold.

\noindent \textbf{RQ2. How is the backlash against DEI affecting software professionals within the software industry?}
While organizational changes are often described in neutral or strategic terms, their impact on people is profoundly personal. The range of emotions that we captured in the posts shows that professionals interpret these shifts through the lens of their own identities, histories, and expectations. The emotional diversity observed reflects deeper tensions about their role as software professionals, belonging, recognition, and the fragility of diversity practices. As software companies recalibrate, software professionals are left to navigate a climate marked by ambiguity and uneven signals about what is safe, supported, or valued.

\noindent \textbf{RQ3. What strategies are currently being used in the software industry to protect and sustain DEI efforts in response to the backlash?}
Across the dataset, we identified forms of resilience that demonstrate how software organizations are attempting to preserve DEI values. Rebranding, strategic framing, and quiet continuity are not just coping mechanisms; they are calculated responses that reveal the complexity of doing DEI work in a contested environment. These strategies suggest that DEI is evolving, not disappearing. By adapting its vocabulary and operational modes, DEI will remain a presence in many software development environments, albeit one that is increasingly shaped by the need to balance principle with pragmatism.

\subsection{Comparing Findings with the Literature}

Recent studies have shown DEI efforts are undergoing significant changes across various sectors, with many organizations reducing the visibility or scope of their initiatives \cite{bryan2025, mcgowan2025navigating, alfonseca2023, palmer2025, wong2025}. These changes include rebranding programs, downsizing DEI-specific roles, and shifting toward broader, less targeted terms like ``belonging'' or ``team culture'' \cite{dave2025, douglas2025returning, einwiller2025addressing}. While much of the current literature has focused on fields like business administration and management, our study extends this work by identifying similar patterns in the software industry and exploring how they align with broader organizational strategies.

Our findings confirm that software companies are undergoing comparable structural transformations. Participants reported the elimination of DEI-specific roles, the withdrawal of public-facing commitments in software products, and a shift in internal messaging within teams. These organizational changes reflect trends already discussed in the literature \cite{einwiller2025addressing, mcgowan2025navigating}, but our study adds a distinct contribution by centering the experiences of software engineers. While prior work has largely investigated institutional shifts or executive strategies, we focused on how these changes are felt by those working directly in software development, discussing the day-to-day impact of these changes on communication, motivation, and perceived support for inclusion.

Another key contribution of our study is the focus on emotional responses. 
While some supported the changes, many
engineers described feelings of anxiety, discouragement, disappointment, and uncertainty as DEI efforts were scaled back. 
This emotional dimension has received limited attention in earlier work, which tends to analyze DEI at the structural or organizational level. Our findings complement recent discussions that the retreat from DEI can have lasting consequences for individual well-being and organizational climate, which may be detrimental to software development environments where collaboration and trust are essential \cite{sitzmann2024don, finn2023representation}. Additionally, while prior literature has documented compliance-oriented adaptations and legal reframing of diversity goals \cite{maizel2024dei, douglas2025returning}, we observed that informal networks and peer-driven practices are continuing to support inclusive values in more discreet ways. 

\subsection{Implications for Research and Practice in Software Engineering}

Our findings have important implications for both research and practice in software engineering. By focusing on how software professionals experience changes in DEI efforts, our study moves the conversation beyond institutional policies and toward the everyday realities of software development. We show that inclusion (or its absence) directly shapes how software engineers engage with their work, their teams, and their organizations. This perspective is essential for understanding the full impact of ongoing shifts in DEI across the software industry.

For researchers, our study suggests the need to move beyond broad discussions of representation and instead explore the lived impacts of DEI changes. This includes how software engineers interpret organizational messaging, navigate uncertainty, and adjust their behaviors in response to shifting norms. We also encourage academics to explore new lines of research, such as how DEI changes affect team collaboration, trust, software quality, and even financial outcomes in 
projects. 

For practitioners, our findings highlight the gap between formal DEI programs and the values held by many software professionals. Even as organizations scale back public commitments, many engineers continue to care about fairness, representation, and belonging. Without support, however, these values can become difficult to act on, leading to silence, discomfort, and reduced psychological safety. For software companies, rather than abandoning DEI, this moment offers an opportunity to strengthen it through everyday practices, for instance by embedding inclusion into team culture through mentoring, onboarding, informal networks, and space for reflection. 

\subsection{Threats to Validity}
Following case study research standards \cite{ralph2020empirical}, we identified and addressed several threats to validity. \textbf{Credibility} was a concern due to the reliance on self-reported experiences, which may be incomplete or selectively recalled. To mitigate this, we included only posts from individuals who clearly identified as software professionals and used direct quotations to ensure that our interpretations were grounded in participants’ own words. \textbf{Transferability} is limited by the non-representative nature of the dataset, which does not allow for statistical generalization. Instead, we expect that our findings can be re-interpreted in other contexts based on the rich contextual descriptions provided. \textbf{Reflexivity} posed a risk given the authors’ personal and academic commitments to inclusion. Regarding this, we explicitly acknowledge that all authors belong to underrepresented groups, but we ensured that our analysis was grounded by consistently returning to the raw data to verify interpretations, including also narratives from participants that are against DEI practices. Finally, regarding \textbf{rigor}, we did not have access to other sources of data to conduct external data triangulation; instead, we relied on a systematic open coding process, comparing patterns across multiple posts and user perspectives.

\section{Conclusions} \label{sec:conclusion}
This study investigated how software professionals are experiencing the ongoing backlash against DEI in the software industry. Our findings highlight three central aspects of this moment: organizational changes affecting DEI programs, the emotional responses of software professionals, and the strategies being developed to support inclusive values.  Our findings suggest that the current moment should not be interpreted as a complete abandonment of DEI within the software industry but as an ongoing recalibration of how inclusion is pursued.

While formal programs may be changing or being reduced, most software professionals in our study demonstrated a strong commitment to DEI values and viewed these principles as important to their work. Looking at software companies, rather than disengaging, several of them are adjusting their approaches, finding new ways to maintain DEI values present in their environments. This ongoing engagement signals that equity, diversity, and inclusion continue to matter within software teams and are being sustained through creativity, resilience, and different forms of resistance.

For future research, we plan to develop a large-scale survey to explore how software professionals across regions and roles perceive DEI in their organizations. We also aim to analyze public statements and corporate communication materials to trace how company-level DEI commitments are evolving. Together, these efforts will deepen our understanding of how the DEI backlash is unfolding and how inclusion is being sustained, transformed, or contested over time.

\section{Data Availability}
The replication package for this study is available at: \url{https://figshare.com/s/b03efc273fd5be2b9227}

\ifCLASSOPTIONcaptionsoff
  \newpage
\fi

\balance
\bibliographystyle{IEEEtran}
\bibliography{bib.bib}

\end{document}